\documentclass[12pt,twoside,a4paper]{article}
\pdfoutput=1

\usepackage{pgfplots}
\pgfplotsset{compat=1.3} 

\usepackage{amssymb}
\usepackage{amsmath}
\usepackage[makeroom]{cancel}
\usepackage{latexsym}
\usepackage{longtable}
\usepackage{epsfig}
\usepackage{graphicx,bbm,psfrag}

\newcommand{\bc}{\begin{center}}
\newcommand{\ec}{\end{center}}
\def\ba#1{\begin{array}{#1}\displaystyle}
\newcommand{\ea}{\end{array}}

\newcommand{\beq}{\begin{equation}}
\newcommand{\eeq}{\end{equation}}
\newcommand{\beqa}{\begin{eqnarray}}
\newcommand{\eeqa}{\end{eqnarray}}

\newcommand{\bi}{\begin{itemize}}
\newcommand{\ei}{\end{itemize}}
\newcommand{\p}{\partial}
\def\lt#1{\left#1}
\def\rt#1{\right#1}

\def\frc#1#2{\frac{#1}{#2}}

\newcommand{\Z}{{\mathbb{Z}}}

\newcommand{\R}{{\mathbb{R}}}

\newcommand{\dd}{d}

\begin{document}

\begin{center}
{\Large {\bf Dynamics of hard rods with initial domain wall state}}

\vspace{1cm}

{\large Benjamin Doyon$^*$ and Herbert Spohn$^\dag$}
\vspace{0.2cm}

{\small\em
$^*$ Department of Mathematics, King's College London, Strand, London WC2R 2LS, U.K.\\
$^\dag$ Physik Department and Zentrum Mathematik, Technische Universit\"at M\"unchen, Boltzmannstrasse 3, 85748 Garching, Germany}
\end{center}

\vspace{1cm}

\noindent Inspired by recent results on the non-equilibrium dynamics of many-body quantum systems, we study the classical hard rod problem in one dimension with initial domain wall condition. Hard rods are an integrable system, in the sense that for each velocity the density of particles is locally conserved. It was proven by Boldrighini, Dobrushin and Sukhov (1983) that on the hydrodynamic space-time  scale, the fluid of hard rods satisfies Euler-type equations 
which comprise all conservation laws.
We provide the general solution to these equations on the line, with an initial condition where the left and right halves are, asymptotically, in different states. The solution is interpreted as being composed of a continuum of contact discontinuities, one for each velocity. This is a classical counterpart of the transport problem solved recently in quantum integrable systems. We provide two independent derivations, one paralleling that in the quantum case, the other using explicitly the microscopic dynamics. Finally, we discuss the Navier-Stokes (viscous) corrections, and study its effect on the broadening of the contact discontinuity and on entropy production.

\vspace{1cm}

{\ }\hfill
\today

\section{Introduction}
\label{sec0}

The study of non-equilibrium, deterministic dynamics of many body systems has received a great amount of attention in recent years. This is particularly true in the quantum context, where investigations into the unitary quantum evolution of extended systems has led to a wealth of new fundamental theoretical principles. For instance, in so-called homogeneous quantum quenches, a coupling parameter of the model is suddenly changed, and the question of thermalization and, for integrable systems, generalized thermalization has been much studied \cite{rev1,rev2,Eisrev,EFreview,VRreview}.

Inhomogeneous states and dynamics are, however, more difficult to approach. Of particular interest is the partitioning protocol for generating non-equilibrium steady states from an initial inhomogeneity, already studied in the 1970ies \cite{spo}. In the partitioning protocol, two halves of an infinite system are independently thermalized (a ``domain wall" initial condition), then connected and let to evolve unitarily for a large time. If the dynamics admits ballistic processes, then at large times, local observables are in a stationary state with nonzero currents, for instance of energy or particles -- this is a non-equilibrium steady state. In the quantum context, Ruelle's work \cite{Ruelle} provided a general mathematical framework, and rigorous results for non-equilibrium currents and some other observables in quantum chains admitting free-fermionic descriptions were obtained \cite{tas,tas2,ah,og,aschp}, see also \cite{AKR08}. The protocol received renewed attention after an exact general solution at one-dimensional quantum critical points was given \cite{BD2012,BD-long}, with numerical verification \cite{Moore12}, experimental evidence \cite{Pierre_et_al}, and a mathematically rigorous proof in the context of algebraic conformal field theory \cite{hl16}. See \cite{lvbd13,SM13,collura1,collura2,Nat-Phys,CKY,D_2014,BDV15,doyonKG,VSDH,BF16,BDirre,Pour,rarefact1,rarefact2,kz16,kar16,LLMM16,LLMM17} and the reviews \cite{BDreview,vm16}. 

The problem of finding exact solutions to the domain-wall initial-value problem in the context of interacting quantum integrable models stayed unsolved for some time, although various conjectures were given \cite{castroint,DeLucaVitiXXZ,Zotos16}. A breakthrough came \cite{cdy,BCDF16} when the powerful ideas of emerging hydrodynamics, whereby a many-body quantum system receives an effective description as a fluid \cite{Qhydro1,Qhydro2,Qhydro3,Qhydro4}, were generalized to the infinite number of conservation laws present in integrable models. In this context, the hydrodynamic theory, called generalized hydrodynamics, has become a fundamental tool in order to study the partitioning protocol and more generally situations with inhomogeneous state or in inhomogeneous force fields, see \cite{dynote,in17,zlp17,bvkm17} and related ideas \cite{dsvc17,s16}.

Despite the large amount of work in the context of quantum integrable models, relatively few results exist in classical integrable models. In this note we study the domain wall problem for an integrable classical dynamics: that of hard rods located on the one-dimensional line. In the usual parlance the model is interacting, thus behaving very differently from an ideal gas and the harmonic chain, for which the exact solution in the case of a domain wall initial state is written down with ease. We mention that a detailed study for Fermi-Pasta-Ulam-type chains with domain wall initial state has been carried out recently \cite{MS16}. But these are non-integrable systems with a dynamical behavior very different from the integrable case discussed here.

Hard rods are segments, say of length $a$, that move freely in one dimension except for elastic collisions whereby they exchange their velocities. Clearly,
velocities are conserved implying an integrable dynamics. Under certain natural conditions, as proven in \cite{dobrods}, at the Eulerian scale, a fluid of hard rods satisfies dynamical equations which account for all local conservation laws. Using these hydrodynamic equations, we provide an exact solution to the domain wall problem, for any initial distribution on the two halves. As is usual in the domain wall problem, this solution is self-similar, and at large times, it is valid for any initial distribution with the same asymptotics on the far left and right. Paralleling the case of quantum integrable models \cite{cdy,BCDF16}, this solution is composed of infinitely many contact singularities, one for each velocity. We provide a derivation inspired by the recent discoveries in quantum integrable models as well as a complementary derivation that takes into account the microscopic dynamics.

One important problem in the study of non-equilibrium systems using hydrodynamic principles is to establish the viscous corrections to the Euler equations and their effects on the dynamics. This problem is still largely unsolved in the quantum context, see however the recent study \cite{zlp17}. For hard rod dynamics, the Navier-Stokes equation  has been derived in \cite{BS97}, offering us the opportunity of studying its effects on the transport problem. We discuss such viscous corrections, and how they affect the broadening of the contact singularities and lead to the production of entropy.  The non-vanishing entropy production signals that, albeit being integrable, hard rods are interacting.

The full relation between the generalized hydrodynamic equations found in quantum integrable systems and the classical hard rod dynamics will be developed in separate works \cite{dyc17,ds17}. Here we simply mention that the hard rod problem can be seen, at the Euler scale, as a generalized hydrodynamic problem, but with a particular choice of the differential scattering phase appearing in the thermodynamic Bethe ansatz \cite{tba,moscaux}.

The paper is organized as follows. In Section \ref{sec1}, we review the hard rods dynamics and Euler scale hydrodynamics. In Section \ref{secdm}, we derive the exact solution to the hydrodynamics with domain wall initial state. In Section \ref{secd}, we discuss diffusive corrections and entropy production. The appendix provides comparison of the exact solution proposed with simulations of the hard rod gas.


\section{Hard rod dynamics}
\label{sec1} 

\subsection{Microscopic dynamics}

Consider a one-dimensional fluid of hard rods. The hard rods have length $a>0$, positions $q_j\in\R$, and momenta $p_j\in\R$. They move freely except for elastic collisions conserving momentum and energy, whereby two hard rods  exchange their momenta upon contact. For convenience we set their mass to $1$, and later will mostly refer to $p_j$ as velocity. In a collision, we label the particles so as to maintain their velocities. Thus particle $j$ moves along a straight line, $\dot{q}_j = p_j$, interspersed with jumps back and forth by $a$ due to collisions. Clearly, such a system with $N$ rods has $N$ conservation laws labeled by velocities. 

Let us first discuss the generalized Gibbs ensembles (GGEs). A GGE is specified by fixing the volume, say the finite interval $[-L,L]$, the density $\rho = N/(2L)$ with $0 < a\rho < 1$, and the single-particle velocity probability density $h(v) \geq 0$ with $\int_\mathbb{R} dv\, h(v) =1$; that is, $h(p_j)dp_j$ is the probability for particle $j$ to have a velocity between $p_j$ and $p_j+d p_j$. Then the canonical ensemble is given by  the probability measure
\begin{equation}\label{0}
Z^{-1} \chi\big(\{|q_i - q_j| \geq a, 1 \leq i < j \leq N \}\big)\prod_{j=1}^N h(p_j)\, \tfrac{1}{N!}dq_jdp_j
\end{equation}
on phase space $([-L,L]\times \mathbb{R})^N$ (where $Z$ is the normalization factor). Here $\chi$ denotes the indicator function of the set specified. Under this equilibrium measure, velocities are independent of the positions and the velocities themselves are independent. The positions are however correlated because of the hard core repulsion. The infinitely extended system is obtained through the limit $N,L \to \infty$ at
fixed $\rho$. The state is fully characterized by $f(v) = \rho h(v)$, and the corresponding equilibrium average will be denoted by $\langle\cdot\rangle_{f}$. In the following we will consider directly the infinite hard rod system.
For the infinite system it is known that in the class of sufficiently regular measures, the only time-stationary measures are given by the above generalized equilibrium 
\cite{dobrods}.

Let us now discuss the dynamics. It is convenient to define densities which will have well-behaved hydrodynamic limits. The particle number density, at time $t=0$, is
\begin{equation}\label{1a}
\mathfrak{n}(x;v) = \sum_{j\in\Z} \delta(x - q_j)\delta(v - p_j).
\end{equation}
That is, $\mathfrak{n}(x;v) dxdv$ is the number of particles in the square $[x,x+dx]\times[v,v+dv]$ of one-particle phase-space. This is a conserved field  thanks to conservation of velocities under collisions, and our somewhat unusual notation emphasizes that $v$ is the label of the conservation law. The time-evolved fields are
\begin{equation}\label{1b}
\mathfrak{n}(x,t;v) = \sum_{j\in\Z} \delta(x - q_j(t))\delta(v - p_j),
\end{equation}
where we recall that the $p_j$'s are independent of $t$ according to our choice of labelling. The particle number density satisfies the local conservation law 
\begin{equation}\label{1}
\partial_t \mathfrak{n}(x,t;v) + \partial_x \mathfrak{j}(x,t;v) = 0
\end{equation}
with current $\mathfrak{j}(x,t;v)$ explicitly given by
\begin{equation}\label{1b2}
\mathfrak{j}(x,t;v) = \sum_{j\in\Z} \dot q_j(t) \,\delta(x - q_j(t))\,\delta(v - p_j),
\end{equation}
where
\beq
	\dot q_j(t) = p_j + a\sum_{j \neq k \in\Z} |p_k-p_j|\big(\chi(p_k<p_j)\delta(q_k(t)-q_j(t)-a) - \chi(p_k>p_j)\delta(q_k(t)-q_j(t)+a)\big)
\eeq
represents the linear motion with jumps at collisions.

The observables $\mathfrak{n}(x,t;v)$ and $\mathfrak{j}(x,t;v)$ are generalized functions on phase space. Integrating against a smooth localized function $g$ on phase space, as $\int dxdv\,g(x,v) \mathfrak{n}(x,t;v)$ or $\int dxdv\,g(x,v) \mathfrak{j}(x,t;v)$, yields functions on phase space. Currents can  be defined also at a given location, $x$, by integrating in time and velocities as 
\begin{equation}\label{1c}
\int_t^{t+\tau}ds\int_\Delta dv\, \mathfrak{j}(x,t;v),
\end{equation}
which counts the signed number of particles with velocities in the interval $\Delta$ crossing the given point $x$ during the time interval $[t,t+\tau]$. The averages of these observables in GGEs are given by 
\begin{equation}\label{1d}
\langle \mathfrak{n}(x,t;v)\rangle_{f} = f(v),                \quad \frac{\langle\mathfrak{j}(x,t;v)\rangle_{f}}{\langle \mathfrak{n}(x,t;v)\rangle_{f}} =v + a(1 - a\rho)^{-1}
\int_\mathbb{R} dw\, (v-w) f(w),
\end{equation}
see \cite{S91} Sec. I.3.3 for more details. 


\subsection{Eulerian hydrodynamics and normal modes}

We now focus on the Eulerian scale. This is the large-distance, long-time scale, which emerges when conserved fields are slowly varying in space, and locally the system is approximately in a generalized equilibrium state. An analysis of the viscous correction (see sub-section \ref{ssectns}) indicates that the Eulerian scale is controlled by the rod length $a$ as follows: it emerges, in some large region, when the space between rods throughout this region, in mean with respect to the local generalized equilibrium, times the mean variation length of the rod density, is much greater than $a^2$. We note that the local equilibrium may be understood in various ways: one may average over space, time, or space-time ``fluid cells'', large enough to give negligible fluctuations, but small enough so that local states are slowly varying; and one may additionally average over realizations of the dynamics with different initial conditions determined by a given distribution, thus allowing the size of fluid cells to be reduced arbitrarily.

We therefore upgrade $f(v)$ to the space-time dependent function $f(x,t;v)$, which fully characterizes the local state at $x,t$. The evolution equation for $f$ is formally obtained by averaging \eqref{1} in a local equilibrium state. To write down the Euler type evolution equation let us first introduce the  local density and velocity
\begin{equation}\label{2}
\rho(x,t) = \int_\mathbb{R} dv f(x,t;v), \quad u(x,t) = \frc1{\rho(x,t)}\int_\mathbb{R} dv\, vf(x,t;v).
\end{equation}
The $x,t$ argument will be omitted if obvious from the context. Using the local averages as provided in \eqref{1d}, the dynamics of $f$ is governed by
\begin{equation}\label{4}
\partial_t f(v) + \partial_x\big(v_{[f]}^\mathrm{eff}(v) f(v)\big) = 0.
\end{equation}
Here
\begin{equation}\label{3}
v_{[f]}^\mathrm{eff}(v)= v + a(1 - a \rho)^{-1} \int_\mathbb{R} dw\,(v-w)f(w)
= v + \frc{a\rho (v-u)}{1 - a \rho},
\end{equation}
which can also be written
\beq
v_{[f]}^\mathrm{eff}(v)=   \frc{v - a \rho u}{1-a\rho}.
\eeq
A tracer particle (in our context also called pulse) with ``bare'' velocity $v$ in fact moves with a corrected, or effective, velocity $v_{[f]}^\mathrm{eff}(v)$ arising from the collisions with other pulses of velocities $w$.  The velocity $v_{[f]}^\mathrm{eff}(v)$ depends explicitly on $v$ and is a functional of $f$. For hard rods the 
dependence is only through the first two moments of $f$ and given by a linear rational function. For integrable quantum models, as for instance the Lieb-Liniger delta-Bose gas, the dependence on $f$ is more complicated \cite{cdy,BCDF16}. As a parenthesis we recall that \eqref{4} holds not only on average. Dobrushin et al. \cite{dobrods} prove a law of large numbers  in the sense that typical initial configurations will follow the hydrodynamic evolution with a probability very close to one. 

Let us briefly deviate from our topic by recalling  the Eulerian hydrodynamic problem in the case of a finite number, say $k$,  conservation laws, $u_\alpha$, $\alpha = 1,...,k$,
governed by
\begin{equation}\label{6}
\partial_t u_\alpha  + \partial_x j_\alpha(\vec{u}) = 0.
\end{equation}
Since currents are functions of densities through the equations of state, this can equivalently be written in quasilinear form as
\begin{equation}\label{7}
\partial_t u_\alpha  + \sum_{\beta =1}^n A_{\alpha\beta}(\vec{u})\partial_x u_\beta = 0, \quad A_{\alpha\beta} =\partial_{u_\beta} j_\alpha.
\end{equation}
 For small deviations from a uniform background, one has to diagonalize $A$. The corresponding normal modes propagate convectively with velocities given by the spectrum of $A$. In many other problems the quasilinear version is the easiest to deal with.

Above, the hydrodynamic variables are the $f(v)$'s. Accordingly, we may bring Eq \eqref{4} into the form \eqref{7}, but the resulting linearized operator $A$ is of a complicated form. Surprisingly, there is a simple transformation diagonalizing $A$. We first introduce the ``free density"
\begin{equation}\label{10}
n(v) = (1-a\rho)^{-1}f(v).
\end{equation}
As to be stressed there is no relation to the microscopic particle density $\mathfrak{n}$ introduced above.
Then 
\begin{equation}\label{11}
\partial_t(1- a \rho) + \partial_x\big(v_{[f]}^\mathrm{eff}(v)(1-a\rho)\big) = - a \big(\partial_t \rho + \partial_x( u \rho )\big) = 0
\end{equation}
by conservation of mass. Hence 
\begin{eqnarray}\label{12}
&&\hspace{-20pt}\partial_t n(v) +v_{[f]} ^\mathrm{eff}(v)\partial_x n(v) \nonumber\\
&&\hspace{-10pt}= (1-a\rho)^{-1}\big(\partial_t f(v) + v_{[f]} ^\mathrm{eff}(v) \partial_x f(v)\big) -  (1-a\rho)^{-2}f(v) \big(\partial_t (1-a\rho) + v_{[f]} ^\mathrm{eff}(v)  \partial_x(1-a\rho)\big)
\nonumber\\
&&\hspace{-10pt} = - (1-a\rho)^{-1}f(v) \partial_x v_{[f]} ^\mathrm{eff}(v) + (1-a\rho)^{-1}f(v) \partial_xv_f ^\mathrm{eff}(v)
 \end{eqnarray}
 and therefore $n(v)$ has vanishing convective derivative,
 \beq\label{neq}
 	\partial_t n(v) +v_{[f]} ^\mathrm{eff}(v)\partial_x n(v) = 0.
 \eeq
The fields $n(v)$ are the normal modes of the hard rod fluid equations, and the convective equation identifies their propagation velocities as $v_{[f]}^\mathrm{eff}(v)$ with $f$ related to $n$ by \eqref{10}.

The free density is closely related to the occupation function of the thermodynamic Bethe ansatz \cite{tba,moscaux} in the quantum problem; see \cite{dyc17,ds17} for the details of this relation.


\section{Domain wall initial state, Euler scale}\label{secdm}

We are interested in solving Eq. \eqref{4} with  the domain wall initial conditions
\begin{equation}\label{5}
f_0(x,v) = f_{-}(v)\,\, \mathrm{for} \,\, x <0,\qquad f_0(x,v) = f_{+}(v)\,\, \mathrm{for} \,\, x >0.
\end{equation}

As before, let us briefly recall the Eulerian hydrodynamic problem in the case of $k$ conservation laws governed by \eqref{6}. The domain wall initial state is invariant under spacial dilations, and the hydrodynamic equation \eqref{6} is invariant under simultaneous dilations of space and time. Therefore, if no spontaneous breaking of this symmetry occurs, the solution is of scaling form,
\begin{equation}\label{8}
u_\alpha(x,t) = \mathsf{u}_\alpha(x/t).
\end{equation}
Setting $\xi = x/t$, one obtains 
\begin{equation}\label{9}
\Big( \sum_{\beta =1}^n \big(\xi\delta_{\alpha\beta} -  A_{\alpha\beta}(\!\vec{\,\mathsf{u}})\big)\Big)\partial_\xi \mathsf{u}_\beta = 0.
\end{equation}
Thus the domain wall solution is related to the eigenvalues of $A$. Since $A$ has discrete spectrum, the actual construction of $\vec{\!\mathsf{\,u}}$, or of the normal modes, requires considerable efforts and we refer to \cite{B13} for a most readable introduction.  As a result of such analysis, generically $\vec{\,\mathsf{u}}$ will have flat pieces interrupted by shocks, contact discontinuities, and rarefaction waves. Relevant for us are the contact discontinuities,
which are defined by the property that the velocity of the jump equals to one of the eigenvalues of $A$.
In our case, the $A$ matrix is infinite-dimensional and has only continuous spectrum. As will be discussed, there are only contact discontinuities. 


\subsection{Exact solution}

We first provide a derivation that closely parallels that given in the quantum context \cite{cdy,BCDF16}. By scale invariance of the initial condition, one expects the solution to Eqs. \eqref{4}, \eqref{5} to be self-similar, of the form 
\begin{equation}\label{13}
f(x,t;v) =\mathsf{f}(\xi;v), \quad n(x,t;v) = \mathsf{n}(\xi;v), \quad v_{[f]} ^\mathrm{eff}(v) = v^\mathrm{eff}(\xi;v).
\end{equation}
Using \eqref{12}, $\mathsf{n}$ satisfies 
\begin{equation}\label{14}
(\xi - v^\mathrm{eff}(\xi;v) \partial_\xi \mathsf{n}(\xi; v) = 0.
\end{equation}
Setting $n_\pm(v) = (1 -a\rho_\pm)^{-1}f_\pm(v)$, this equation together with the boundary condition \eqref{5} is solved by
\begin{equation}\label{15}
\mathsf{n}(\xi;v) = n_-(v)\chi(v>v_*(\xi)) + n_+(v)\chi(v<v_*(\xi)),
\end{equation}
where $v_*$ is defined implicitly through 
\begin{equation}\label{16}
v^\mathrm{eff}(\xi;v_*) = \xi.
\end{equation}

Let us first make sure that \eqref{16} has a unique solution. Introducing 
\begin{equation}\label{16a}
\kappa(v) = \int_{v}^\infty dw\,  n_-(w) + \int^{v}_{-\infty} dw\,  n_+(w), \quad
\nu(v) = \int_{v}^\infty dw\, w n_-(w) + \int^{v}_{-\infty} dw \,w n_+(w)
\end{equation}
and 
using $v^\mathrm{eff}(v) = v + a \rho (1-a\rho)^{-1}( v -u) $, one obtains 
\begin{eqnarray}\label{17}
&&\hspace{-20pt} \xi = v_* - a \int _\mathbb{R}dv (v - v_*)\mathsf{n}(\xi;v)  \nonumber\\
&&\hspace{-10pt}=  v_* - a\Big( \int_{v_*}^\infty dv \,v n_-(v) + \int^{v_*}_{-\infty} dv \,v n_+(v)\Big) + a v_*\Big( \int_{v_*}^\infty dv  \,n_-(v) + \int^{v_*}_{-\infty} dv \, n_+(v)\Big) \nonumber\\
&&\hspace{-10pt} = v_* - a \nu(v_*) + a v_*\kappa(v_*) = g(v_*).
 \end{eqnarray}
Then 
\begin{equation}\label{18}
g' = 1 -a \nu' +a\kappa +av\kappa' = 1+a \kappa.
\end{equation}
Hence $g'>0$, $g$ is invertible, and $v_* = g^{-1}(\xi)$. For large $\xi$, one has $v_* = \xi$. Thus the solution \eqref{15} indeed satisfies the boundary condition \eqref{5}.

The final step is to switch back to the original field.  Observing that $(1 - a\rho)^{-1} = 1+ a\int_\mathbb{R}dv\,\mathsf{n}(v)$ one concludes
\begin{equation}\label{18a}
(1 - a\rho (\xi))^{-1} =  1+a \kappa (v_*) = g'(v_*).
\end{equation}
Hence
\begin{equation}\label{19}
\mathsf{f}(\xi,v) = \frac{1}{g'(g^{-1}(\xi))}\Big( \frac{1}{1- a\rho_-}f_-(v) \chi(\xi<g(v)) +  \frac{1}{1- a\rho_+}f_+(v) \chi(\xi >g(v)\Big).
\end{equation}

Reading the identity \eqref{16}, the solution \eqref{15} is naturally interpreted as a family of contact discontinuities. This is the same type of solution as was found in \cite{cdy,BCDF16} in the case of generalized hydrodynamics for quantum systems. See Appendix \ref{app} for comparisons between this exact solution and numerical simulations of the hard rod gas. Note that in the quantum context, unicity of the solution to \eqref{16} was only verified numerically. Here, since the effective velocity has a simpler form, unicity could be proved. The explicit solution \eqref{19} also is particular to the hard rod problem, but a generalization to other differential scattering phases can be obtained as integral equations \cite{dys17}.


\subsection{Microscopic derivation}\label{ssectmicro}
The domain wall solution \eqref{19} can be guessed by considering the microscopic hard rod dynamics and appealing to a 
suitable law of large numbers. 
So let us see how this works. At $t=0$ we shrink $a$ to zero. Then $f_0(x;v)$ goes over to
$\tilde{f}_0(y;v) = (1-a\rho_-)^{-1} f_-(v) \chi(y<0) + (1-a\rho_+)^{-1} f_+(v) \chi(y>0)$, which evolves under the free dynamics to
\begin{equation}\label{20}
\tilde{f}_0(y,t;v) = (1-a\rho_-)^{-1} f_-(v) \chi(y-vt<0) + (1-a\rho_+)^{-1} f_+(v) \chi(y-vt>0).
\end{equation}
We have to compute how a small volume element at $y$ with velocity $v$ is shifted to $x$
upon expanding the hard rods back to size $a$. This shift has two contributions. Firstly, under the free dynamics the initial density jump
at $0$ is shifted to $vt$. Upon expanding back to the original hard rod length, for $vt<y$ the particles in $[vt,y]$ yield the shift $ (1-a\rho_+)^{-1}\rho_+(y-vt)a$, while for
 $vt>y$ the particles in $[y,vt]$ yield the shift $ -(1-a\rho_-)^{-1}\rho_-(vt - y)a$. Hence in sum
 \begin{equation}\label{21}
a(y-vt)\big((1-a\rho_-)^{-1}\rho_- \chi(y < vt)+ (1-a\rho_+)^{-1}\rho_+ \chi(y>vt)\big).
\end{equation}

A second contribution comes from other particles, say with velocity $w$, crossing the positions $y-v(t-s)$ at times $0 \leq s \leq t$, which causes a shift $-a$ upon crossing from 
left to right and shift $a$  from right to left. Thus we have to compute $a$ times the total number of signed crossings as
 \begin{equation}\label{22}
a\mathfrak{j}(y,t;v) = -a\int_\mathbb{R}dw\Big( \chi(w>v) \int_{y-wt}^{y-vt}dy'\tilde{f}_0(y',0;w) - \chi(w<v) \int_{y-vt}^{y-wt}dy'\tilde{f}_0(y',0;w)\Big).
\end{equation}
Using that $\tilde{f}_0$ is a step function one divides into the sets $\{ y -wt <0\}$   and $\{ y -wt > 0\}$, respectively  $\{ y -vt <0\}$   and $\{ y -vt > 0\}$.
Working out the integrals, 
\begin{eqnarray}\label{23}
&& \hspace{0pt} a \mathfrak{j}(y,t;v) = a \big( - \nu(y/t) + (y/t)\kappa(y/t)\big) \nonumber\\ 
&& \hspace{60pt}- a(y-vt)\big((1-a\rho_-)^{-1}\rho_- \chi(y < vt)+ (1-a\rho_+)^{-1}\rho_+ \chi(y>vt)\big).
\end{eqnarray}
The second summand precisely cancels the contribution from  Eq. \eqref{21}. We arrived at the expected ballistic scaling and may set $t=1$. Then
the point $x$ is shifted to $x = y + a(-\nu(y) +y \kappa(y)) =g(y)$ with volume element $dx = g'(y) dy$. Inserting in \eqref{20} we arrive at the domain wall solution \eqref{19}.

A re-statement of the above derivation is as follows. Consider the coordinate $y(x,t) = \int_{-x_0}^x \dd x'\,(1-a\rho(x',t)) + x_0$. This is the free length from $x_0$ to $x$, after the rods have been shrunk to points. In terms of this coordinate, the solution to the domain-wall problem becomes quite transparent. We choose $x_0$ far enough on the left so that it lies within the left bath up to large enough times. Then, integrating \eqref{18a} over $x$ at $t$ fixed, we have
\beq
	g^{-1}(x/t) - g^{-1}(x_0/t) = (y(x,t)-x_0)/t.
\eeq
Since $v_*$, as defined in \eqref{16}, is equal to $g^{-1}(\xi)$, and since $v_*\to\xi$ as $\xi\to-\infty$, then we may use $g^{-1}(x_0/t)= x_0/t$ for $|x_0|$ large enough. Therefore,
\beq
	g^{-1}(\xi) = y(\xi t)/t.
\eeq
That is, the function $g^{-1}(\xi)$, which determines the velocity at which a jump occurs in the solution \eqref{15} for $n(\xi;v)$, is nothing else but the ray in the $y$ coordinate. Such geometric ideas are at the basis of the viewpoint provided in \cite{dys17}.


\section{Diffusive corrections}\label{secd}
\subsection{Navier-Stokes correction}\label{ssectns}
On the Eulerian time scale, for fixed $v$, the domain wall solution has a jump discontinuity at $g(v)$. Physically one expects a broadening of the step, which analytically should be accessible through  a suitable Navier-Stokes correction of the Euler equations \eqref{4}. To figure out such a correction 
we follow the standard recipe for computing transport coefficients in equilibrium, as applied to hard rods  in \cite{LPS68, S82}.

For the velocity distribution we recall that $f(v) = \rho h(v)$ with $\int dv\,h(v) = 1$, and we set $\int dv \,h(v)v = 0$, for convenience. On large scales the static fields are $\delta$-correlated in $x$ and  correlated in velocity space as
\begin{equation}\label{24}
 \langle \mathfrak{n}(x,0;v)\mathfrak{n}(x',0;v')\rangle_{\rho h} -  \langle \mathfrak{n}(0,0;v)\rangle_{\rho h}
 \langle\mathfrak{n}(0,0;v')\rangle_{\rho h}= \rho\, \delta(x-x') C(v,v')
\end{equation}
with
\begin{equation}\label{25}
C(v,v') = \rho\big(\delta(v-v')h(v) + a\rho (a\rho - 2) h(v)h(v')\big).
\end{equation}
The second summand results from the integral over the static density-density correlation.
Linearizing Eq. \eqref{4} as $\rho h(v) +f(v)$ yields the linearized Euler operator with integral kernel
\begin{equation}\label{26}
A(v,v') = (1-a\rho)^{-1}\delta(v-v')v +  \rho a(1-a\rho)^{-2}vh(v) - a\rho (1-a\rho)^{-1}v'h(v).
\end{equation}
Hence
\begin{equation}\label{27}
(AC)(v,v') = \rho(1-a\rho)^{-1}\big(\delta(v-v')vh(v) -  a\rho  (v+v')h(v)h(v')\big).
\end{equation}
Clearly $AC=CA^{\mathrm{T}}$, in agreement with linear fluctuating hydrodynamics \cite{S91}.

In our context, the most interesting feature is the total current correlation function, which turns out to be given by
\begin{eqnarray}\label{28}
&&\hspace{-28pt} \int dx \big(\langle \mathfrak{j}(x,t;v)\mathfrak{j}(0,0;v')\rangle_{\rho h} - 
\langle \mathfrak{j}(0,0;v)\rangle_{\rho h}\langle\mathfrak{j}(0,0;v')\rangle_{\rho h}\big)\nonumber\\
&&\hspace{10pt} = \delta(t) (a\rho)^2 (1-a\rho)^{-1}\big(\delta(v-v') r(v)h(v)  - |v-v'|h(v)h(v')\big)\nonumber\\[1 ex]
&&\hspace{20pt}+\rho(1-a\rho)^{-2}\big(\delta(v - v')v^2h(v) - a\rho (v^2 + v'^2) h(v)h(v')  + (a\rho)^2 d_2h(v)h(v') \big),\nonumber\\
&&\hspace{20pt}
r(v) = \int dw h(w)|w-v|, \quad d_2 = \int dw h(w)w^2 .
\end{eqnarray}
The constant term is the Drude weight. As valid in great generality it equals $ACA^{\mathrm{T}}$, compare with  the discussion in  \cite{MS15}.
To obtain the diffusion matrix, $D$, we use the Green-Kubo formula according to which $ DC$ equals the integral over the total current correlation with the Drude weight subtracted.
Hence
\begin{equation}\label{29}
(DC)(v,v') = (a\rho)^2 (1-a\rho)^{-1}\big(\delta(v-v') r(v)h(v) - |v-v'|h(v)h(v')\big),
\end{equation}
which by using \eqref{25} implies
\begin{equation}\label{30}
D(v,v') = a(a\rho) (1-a\rho)^{-1}\big(\delta(v-v') r(v) - h(v)|v-v'|\big).\medskip
\end{equation}
\textit{Remark}. In Eq. I.7.67 of \cite{S91} it is claimed that 
\begin{equation}\label{30a}
D(v,v') = a(a\rho) (1-a\rho)^{-1}\big(\delta(v-v') r(v) - h(v')|v-v'|\big),
\end{equation}
which is incorrect. This oversight leads  to a form of the nonlinear Navier-Stokes equations slightly different from the correct version given below 
in \eqref{33}. Thereby a longstanding discrepancy as regards to \cite{BS97} is \medskip resolved. 

The operator $-D$ is the forward generator of a jump process on $\mathbb{R}$ with jump rate $ |v-v'|h(v')$ for the transition from $v$ to $v'$. The invariant measure is $h(v)dv$ and the rates satisfy detailed balance with respect to $h$. In particular, the quadratic form of $D^\mathrm{T}$ in $L^2(\mathbb{R}, h(v)dv) = \mathcal{H}$ reads
\begin{eqnarray}\label{31}
&&\hspace{-40pt}\langle \phi, D^\mathrm{T}\psi\rangle_\mathcal{H} = \int dvdv' h(v)\phi(v)D(v',v)\psi(v') = \nonumber\\
&&\hspace{20pt}\tfrac{1}{2} a(a\rho) (1-a\rho)^{-1}\int dv dv'|v - v'|h(v)h(v') (\phi(v) - \phi(v'))(\psi(v) - \psi(v')). 
\end{eqnarray}
Hence $D^\mathrm{T} = (D^\mathrm{T})^*$ in $\mathcal{H}$ and the eigenvalues of $D^\mathrm{T}$, thus also of $D$, are nonnegative. 
  The space-time correlator is well approximated by 
 \begin{eqnarray}\label{32}
&&\hspace{0pt} \int_\mathbb{R} dx \mathrm{e}^{-\mathrm{i}kx}\big(\langle \mathfrak{n}(x,t;v)\mathfrak{n}(0,0;v')\rangle_{f}
 -  \langle \mathfrak{n}(0,0;v)\rangle_{f}
 \langle\mathfrak{n}(0,0;v')\rangle_{f}\big) \nonumber\\
 &&\hspace{70pt} \simeq \big(\exp\big[-\mathrm{i}kAt - \tfrac{1}{2} k^2D|t|\big]C\big)(v,v').
 \end{eqnarray}
An explicit error bound can be found in \cite{S82}. 

On this basis one guesses that the Navier-Stokes correction is given by
\begin{equation}\label{33}
\partial_t f(v) + \partial_x\big(v_{[f]} ^\mathrm{eff}(v)f(v)\big) = \p_x \mathcal{N}_{[f]}(v),
 \end{equation}
 where the dissipative term is
\beq\label{diss}
	\mathcal{N}_{[f]}(v) = \tfrac{1}{2} a^2(1-a \rho)^{-1}\int_\mathbb{R}dw |v-w|\big(f(w)\partial_x f(v) - f(v)\partial_x f(w)\big).
\eeq
A mathematical proof is established by Boldrighini and Suhov \cite{BS97}. Note that linearizing \eqref{33} as $\rho h(v) + f(v)$ one obtains indeed \eqref{32}. As a further simple consistency check, according to \eqref{33} one has
\begin{equation}\label{34}
\partial_t \rho  + \partial_x (u \rho) = 0,
 \end{equation}
which must hold, since it is a property for each microscopic trajectory.
 
\subsection{Entropy production}
For usual  hydrodynamics with a few conservation laws, there is a balance equation for the entropy which consists of a local entropy flow and 
a non-negative entropy production proportional to the transport coefficients. In fact, on the Euler scale,  even starting with smooth initial data, shocks may develop spontaneously at some finite time and entropy is produced at shocks and possibly other
non-smooth parts of the solution.  For hard rods, most likely such a mechanism does not work. One evidence comes from \cite{dobrods}, where the uniqueness of solutions for \eqref{4} is established without an entropy condition, which would be required in the case of a finite number of conservation laws. Another argument in support of the absence of the spontaneous formation of stable entropy-producing structures comes from the presence of a continuous set of conservation laws \cite{cdy}. These offer an escape route permitting, on the Euler scale, entropy to be conserved at all times (except possibly for discrete sets of space-time points where infinitesimal amounts of entropy may be produced). For instance, in the domain-wall problem, where the eigenvalue equation \eqref{14} must be solved, the set of conservation laws provides a continuous infinity of eigenvectors, allowing for continuous fluid states\footnote{Since a fluid state is a function of $v$, continuity must be taken in an appropriate topology, which should be that induced by an appropriate completion of the set of local observables.} to join any two reservoirs (an equivalent expression is that a continuum of contact singularities form -- no entropy is produced at such singularities). This is in sharp contrast to the finite case, where usually no continuous solution exists and shocks must form. Thus, in the hard rod hydrodynamics, we expect no entropy-producing structures to spontaneously develop, and entropy production to result only from dissipation.

%

The thermodynamic entropy has two contributions. Firstly, there is the positional part including the hard core repulsion with entropy 
$s_\mathrm{hr}(\rho) = \rho \log(\rho^{-1}(1 - a \rho))$ \cite{LPS68}.
 Secondly, since velocities are independent, the momentum part yields the entropy
\begin{equation}\label{36}
- \rho \int_\mathbb{R} dv\, h(v) \log  h(v) = -  \int_\mathbb{R} dv \,\rho h(v) \log  (\rho h(v) ) + \rho \log \rho. 
 \end{equation}
Hence the local entropy equals
\begin{equation}\label{37}
s(x,t) = -  \int_\mathbb{R} dv f(x,t;v) \log f(x,t;v) + \rho(x,t) \log (1 -a \rho (x,t)). 
\end{equation} 
Note that this can be rewritten as
\beq
	s(x,t) = -(1-a\rho(x,t)) \int_\mathbb{R} dv\, n(x,t;v)\log n(x,t;v)
\eeq
in terms of the free density \eqref{10},  which provides a natural interpretation: if the rods were shrunk to points, the resulting density of free particles yields the entropy.

We remark that this entropy function is different from the so-called Yang-Yang entropy function occurring in generalized equilibrium states of fermionic quantum systems \cite{tba,moscaux}, even if both cases are described by similar equations of states. In the quantum case, there would be an additional term counting the hole configurations. This, of course, points to the fact that the entropy function is an additional characterization of the fluid, not a property of the Euler scale. Different fluids with the same Euler equations may have different entropy functions, with positive entropy production under different viscosity terms.

In Eulerian hydrodynamics, because of \eqref{neq}, the free density $n(x,t;v)$ has zero convective derivative, hence so does any function of it.  Further, by \eqref{11} the function $(1-a\rho(x,t))$ is a conserved density. As a consequence, $(1-a\rho(x,t)) n(x,t;v)\log n(x,t;v)$  is also a conserved density, with the same current. Integrating over $v$, the Eulerian contribution to $\p_t s$ can be immediately evaluated. Taking into account dissipation, one therefore obtains
\beq\label{39}
	\p_t s = \p_x \Big(\int_\mathbb{R} dv\, v_{[f]} ^\mathrm{eff} f\log f -\rho u \log (1 - a\rho)\Big) 
	-  \int_\mathbb{R} dv \,(\log f)\, \partial_x\mathcal{N}_{[f]},
\eeq
where the dissipative term has been simplified using \eqref{34} and $\int_\mathbb{R} dv \,\mathcal{N}_{[f]}(v) = 0$.

We rewrite this as
 \begin{equation}\label{41}
 - \int_\mathbb{R} dv \,(\log f) \,\partial_x\mathcal{N}_{[f]} = - \partial_x \int_\mathbb{R} dv \,\mathcal{N}_{[f]} \log f
 + \int_\mathbb{R} dv\,f^{-1}(\partial_x f)\mathcal{N}_{[f]} .
  \end{equation} 
The first summand is the additional flow term  
 \begin{equation}\label{42}
  \int_\mathbb{R} dv\, \mathcal{N}_{[f]} \log f = \tfrac{1}{2} a^2 (1 -a\rho)^{-1} \int_{\mathbb{R} ^2}dw dv\, |w-v| \log f(v) \big( f(w) \partial_x f(v) - f(v)\partial_xf(w)\big).
\end{equation} 
The second summand is the entropy production
\begin{eqnarray}\label{43}
&& \hspace{-20pt} \sigma_{[f]} =  \int_\mathbb{R} dv\,f^{-1}(\partial_x f)\mathcal{N}_{[f]} \\
&& \hspace{-20pt}= \tfrac{1}{2} a^2 (1 -a\rho)^{-1} \int_{\mathbb{R}^2} dw dv\, |w-v|f(v)^{-1} \partial_x f(v)
 \big( f(w) \partial_x f(v) - f(v)\partial_x f(w)\big) \nonumber\\
&& \hspace{-20pt}= \tfrac{1}{4} a^2 (1 -a\rho)^{-1} \int_{\mathbb{R}^2} dw dv\, |w-v| \Big(\sqrt{f(w)/f(v)}\,\partial_x f(v)-
 \sqrt{f(v)/f(w)}\,\partial_x f(w)\Big)^2 \geq 0.\nonumber
 \end{eqnarray}
  
 Thus the entropy balance reads
  \begin{equation}\label{44}
  \partial_t s +\partial_x j_\mathrm{s} = \sigma_{[f]}
\end{equation} 
with the entropy production defined in Eq. \eqref{43} and the entropy flow
\begin{eqnarray}\label{45}
&& \hspace{-20pt} j_\mathrm{s} = -  \int_\mathbb{R} dv \,v_{[f]} ^\mathrm{eff} f\log f + u\rho \log (1 - a\rho)\\
&& \hspace{20pt} - \tfrac{1}{4} a^2 (1 -a\rho)^{-1} \int_{\mathbb{R} ^2}dw dv\, |w-v| \log (f(v)/f(w)) \big( f(w) \partial_x f(v) - f(v)\partial_xf(w)\big).\nonumber
\end{eqnarray} 
The velocity part of the entropy is transported by the effective velocity $ v_{[f]} ^\mathrm{eff}$ and the spatial part of the entropy by the mean velocity $u$. In addition, there is a contribution proportional to the spatial gradient, which is comparable to the entropy flow for compressible Navier-Stokes containing a term proportional to the temperature gradient. 
\subsection{Domain wall including dissipation}
It seems rather unlikely to still find an explicit domain wall solution once the dissipative term is included. But on a more qualitative level the dynamical
 behavior can be easily guessed. The discontinuity at $\xi = g(v)$ will be smeared out diffusively to become an error function. The diffusion is correlated
 and nonlinear. But if one just assumes constant diffusion a simple picture emerges. Let us consider the density $\rho(x,t) = \rho_\mathrm{s}(x/t)$
 with $\rho_\mathrm{s}$ the $v$-integral of Eq. \eqref{19}. Dissipation generates then a convolution of this profile with a Gaussian of width $\sqrt{t}$.
 If $h_\pm$ are smooth, then $\rho(x,t)$ is smooth because of the $v$-integration. For long times the convolution will hardly change the profile.
 In fact the most visible change should be detected at short times when $\rho$ has still large gradients. The same observation applies to  
 the other physical fields, such as velocity
 and energy.     

\section{Conclusions}

Inspired by recent developments in quantum many-body integrable models, we provided the exact solution to the general domain-wall initial value problem for the hard rod gas on the Euler scale. The solution was obtained in two independent ways, one following the solution strategy of \cite{cdy,BCDF16}, the other based on the explicit microscopic dynamics of the hard rod gas. We then studied viscosity and entropy, establishing the exact entropy current equation with a source controlled by viscosity, and showed positivity of entropy production.

It would be interesting to further understand the effects of viscosity on the domain-wall solution. Hard rod hydrodynamics is of course applicable beyond the exact self-similar limit of the solution to the domain initial value problem, and it would be instructive to study further its range of applicability. Finally, it would be very interesting to study the large deviation theory of the fluctuating currents, perhaps via an extension of the macroscopic fluctuation theory \cite{mft}.

\appendix

\section{Molecular dynamics simulations}\label{app}

In this appendix, we compare the exact solution \eqref{19} with numerical simulations of the hard rod dynamics. The domain wall initial state is taken to be at uniform temperature with a jump in the density. Another popular choice is the converse, uniform density and jump in the temperature, for which we expect similar results. Our simulation is performed at infinite volume.
Initially, besides the discontinuity at the origin,  the density jumps to zero at some point to the right and also to the left. Thus there are two depletion zones moving inwards. The exact solution looses validity once it merges with the depletion zones.
For the exact solution, we use the formulae $\rho = (1-1/g'(v_*))/a$ and $\rho u = \nu(v_*)/g'(v_*) = (v_*-g(v_*)/g'(v_*))/a$ with $v_* = g^{-1}(x/t)$.

\begin{figure}
\bc
\begin{tikzpicture}[scale=0.8] \begin{axis}[
title=(a), xlabel={$x$}, ylabel={$\rho(x)$},
] \addplot[blue] table {dm2-0-density.dat}; 
\end{axis} \end{tikzpicture} \begin{tikzpicture}[scale=0.8] \begin{axis}[
title=(b), xlabel={$x$},
] \addplot[blue] table {dm2-1-density.dat};
\addplot[red] table {dm2-1-solution.dat}; \end{axis} \end{tikzpicture}

\begin{tikzpicture}[scale=0.8] \begin{axis}[
title=(c), xlabel={$x$}, ylabel={$\rho(x)$},
] \addplot[blue] table {dm2-2-density.dat};
\addplot[red] table {dm2-2-solution.dat}; \end{axis} \end{tikzpicture}
\begin{tikzpicture}[scale=0.8] \begin{axis}[
title=(d), xlabel={$x$},
] \addplot[blue] table {dm2-4-density.dat};
\addplot[red] table {dm2-4-solution.dat}; \end{axis} \end{tikzpicture}

\begin{tikzpicture}[scale=0.8] \begin{axis}[
title=(e), xlabel={$x$}, ylabel={$\rho(x)$},
] \addplot[blue] table {dm2-6-density.dat};
\addplot[red] table {dm2-6-solution.dat}; \end{axis} \end{tikzpicture}
\begin{tikzpicture}[scale=0.8] \begin{axis}[
title=(f), xlabel={$x$},
] \addplot[blue] table {dm2-8-density.dat};
\addplot[red] table {dm2-8-solution.dat}; \end{axis} \end{tikzpicture}
\ec
\caption{Density at times (a) $t=0$, (b) $t=0.5$, (c) $t=1$, (d) $t=2$, (e) $t=3$, (f) $t=4$, rod length $a = 0.001$, initial data corresponding to \eqref{59}. Simulation data in blue, exact solution in red.}
\label{f2}
\end{figure}
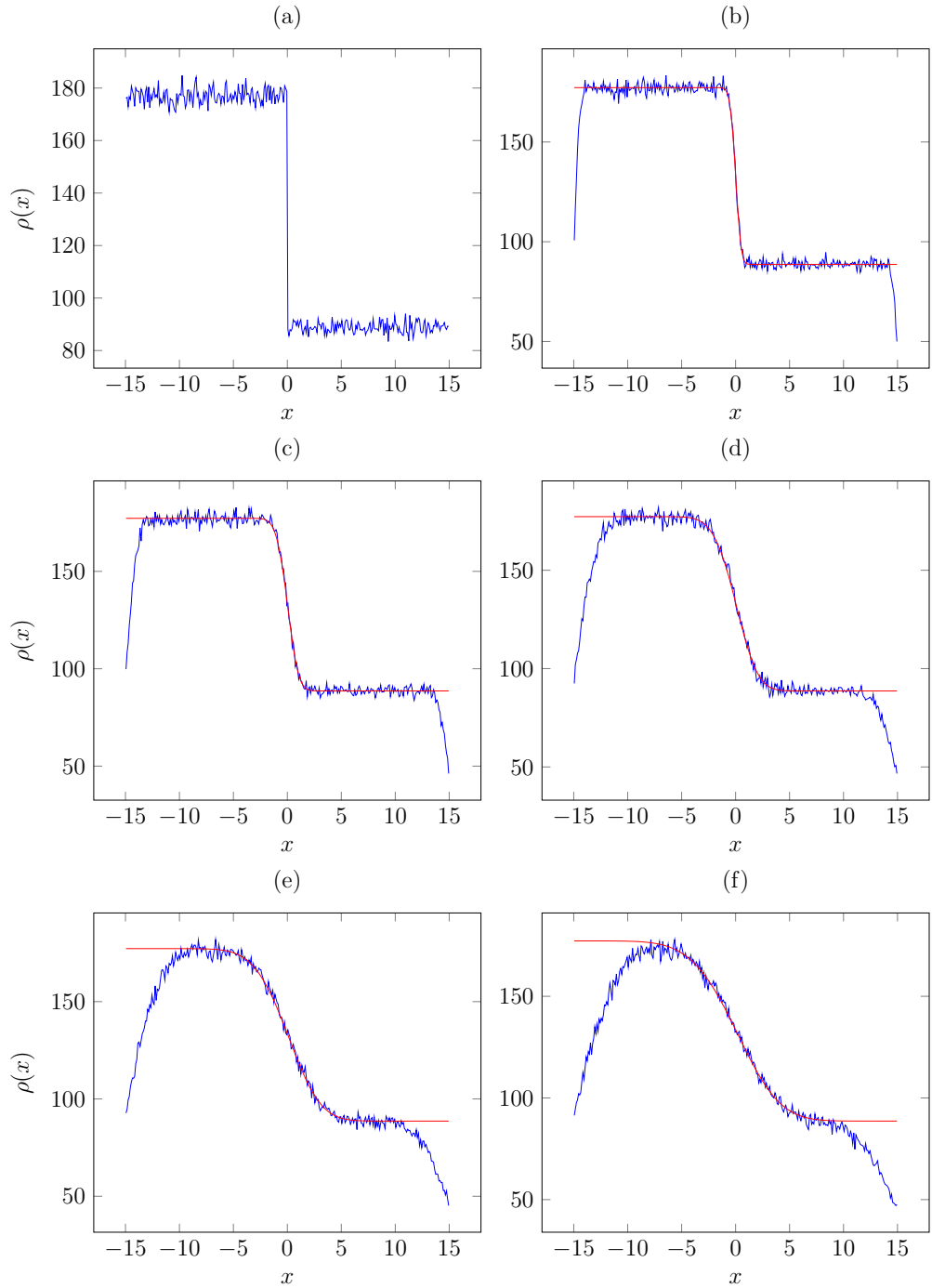

\begin{figure}
\bc
\begin{tikzpicture}[scale=0.8] \begin{axis}[
title=(a), xlabel={$x$}, ylabel={$\rho(x)u(x)$},
] \addplot[blue] table {dm2-2-current.dat};
\addplot[red] table {dm2-2-solutioncur.dat}; \end{axis} \end{tikzpicture}
\begin{tikzpicture}[scale=0.8] \begin{axis}[
title=(b), xlabel={$x$},
] \addplot[blue] table {dm2-4-current.dat};
\addplot[red] table {dm2-4-solutioncur.dat}; \end{axis} \end{tikzpicture}

\begin{tikzpicture}[scale=0.8] \begin{axis}[
title=(c), xlabel={$x$}, ylabel={$\rho(x)u(x)$},
] \addplot[blue] table {dm2-6-current.dat};
\addplot[red] table {dm2-6-solutioncur.dat}; \end{axis} \end{tikzpicture}
\begin{tikzpicture}[scale=0.8] \begin{axis}[
title=(d), xlabel={$x$},
] \addplot[blue] table {dm2-8-current.dat};
\addplot[red] table {dm2-8-solutioncur.dat}; \end{axis} \end{tikzpicture}
\ec
\caption{Current at times (a) $t=1$, (b) $t=2$, (c) $t=3$, (d) $t=4$. , rod length $a = 0.001$, initial data corresponding to \eqref{59}. Simulation data in blue, exact solution in red. }
\label{f2c}
\end{figure}
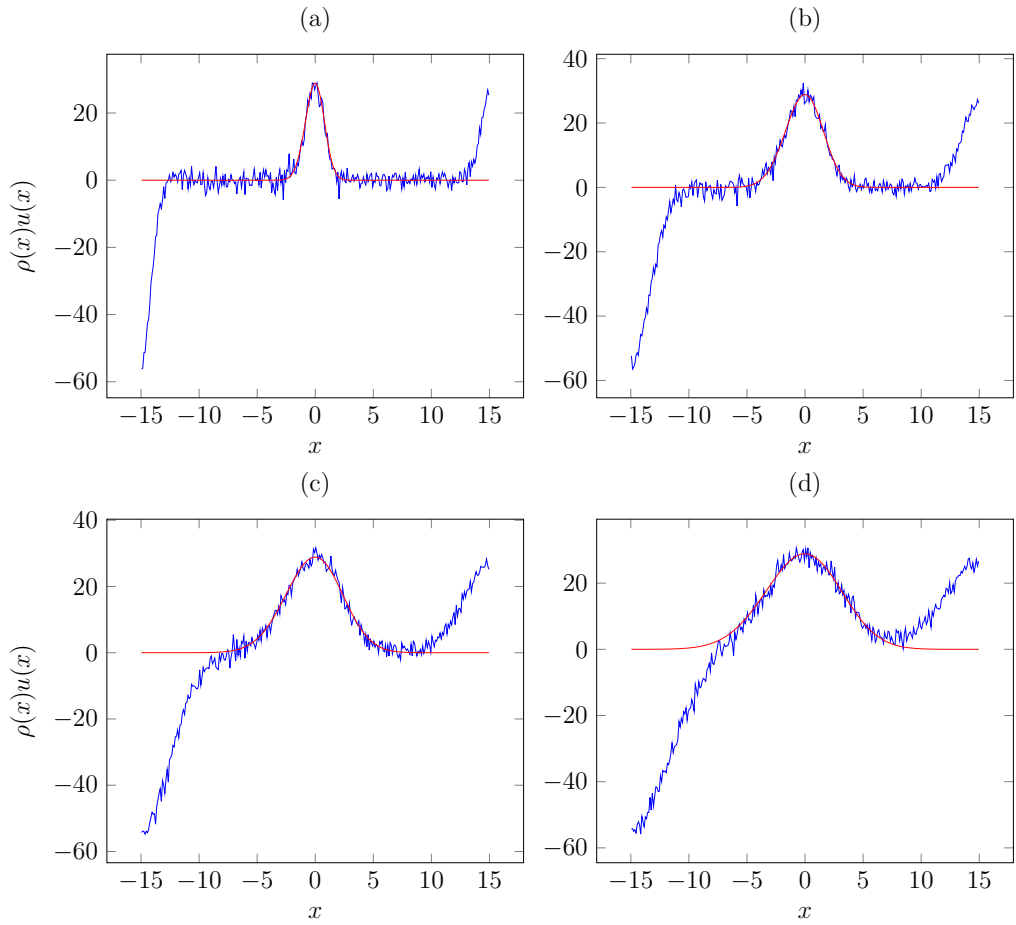

\begin{figure}
\bc
\begin{tikzpicture}[scale=0.8] \begin{axis}[
title=(a), xlabel={$x$}, ylabel={$\rho(x)$},
] \addplot[blue] table {dm3-1-density.dat};
\addplot[red] table {dm3-1-solution.dat}; \end{axis} \end{tikzpicture}
\begin{tikzpicture}[scale=0.8] \begin{axis}[
title=(b), xlabel={$x$},
] \addplot[blue] table {dm3-2-density.dat};
\addplot[red] table {dm3-2-solution.dat}; \end{axis} \end{tikzpicture}

\begin{tikzpicture}[scale=0.8] \begin{axis}[
title=(c), xlabel={$x$}, ylabel={$\rho(x)$},
] \addplot[blue] table {dm3-4-density.dat};
\addplot[red] table {dm3-4-solution.dat}; \end{axis} \end{tikzpicture}
\begin{tikzpicture}[scale=0.8] \begin{axis}[
title=(d), xlabel={$x$},
] \addplot[blue] table {dm3-6-density.dat};
\addplot[red] table {dm3-6-solution.dat}; \end{axis} \end{tikzpicture}
\ec
\caption{Density at times (a) $t=0.5$, (b) $t=1$, (c) $t=2$, (d) $t=3$, rod length $a = 0.001$, initial data corresponding to \eqref{60}. Simulation data in blue, exact solution in red. }
\label{f3}
\end{figure}
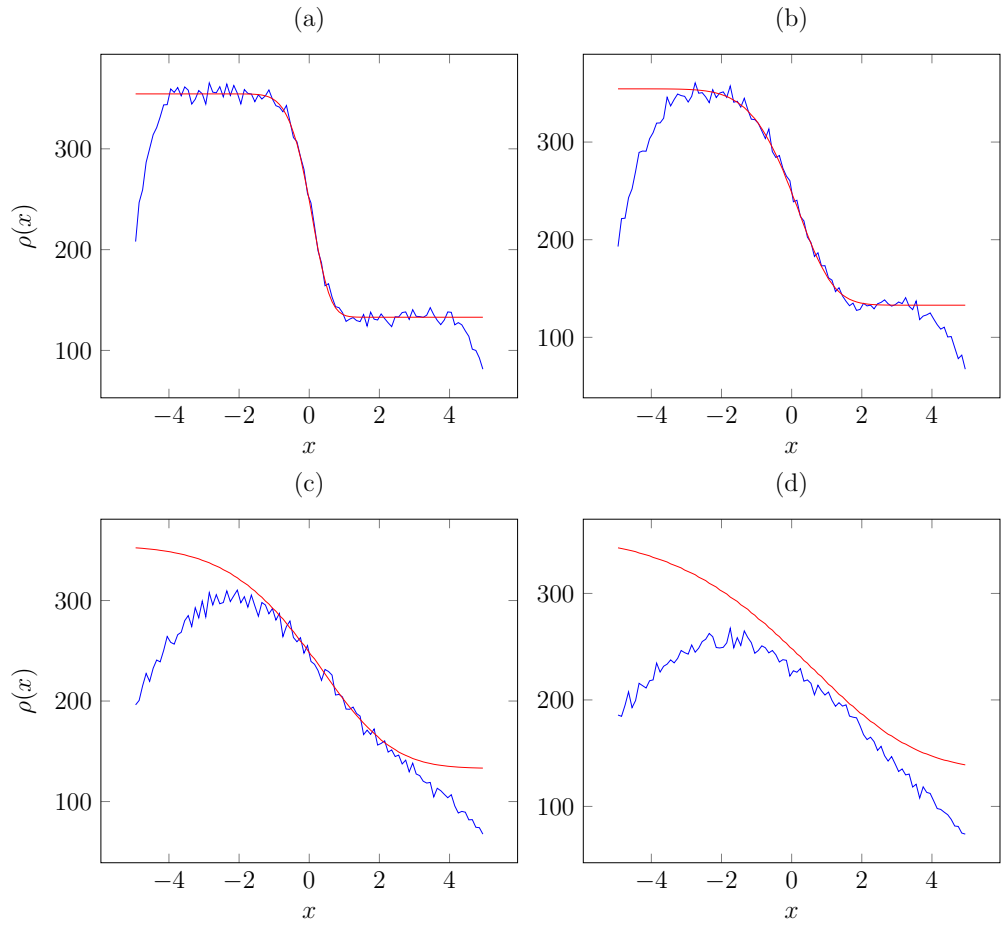

\begin{figure}
\bc
\begin{tikzpicture}[scale=0.8] \begin{axis}[
title=(a), xlabel={$x$}, ylabel={$\rho$},
] \addplot[blue] table {dm5-0-density.dat}; \end{axis} \end{tikzpicture}
\begin{tikzpicture}[scale=0.8] \begin{axis}[
title=(b), xlabel={$x$}, 
] \addplot[blue] table {dm5-2-density.dat};
\addplot[red] table {dm5-2-solution.dat}; \end{axis} \end{tikzpicture}

\begin{tikzpicture}[scale=0.8] \begin{axis}[
title=(c), xlabel={$x$}, ylabel={$\rho$},
] \addplot[blue] table {dm5-4-density.dat};
\addplot[red] table {dm5-4-solution.dat}; \end{axis} \end{tikzpicture}
\begin{tikzpicture}[scale=0.8] \begin{axis}[
title=(d),  xlabel={$x$}, 
] \addplot[blue] table {dm5-6-density.dat};
\addplot[red] table {dm5-6-solution.dat}; \end{axis} \end{tikzpicture}
\ec
\caption{Density at times (a) $t=0$, (b) $t=1$, (c) $t=2$, (d) $t=3$, rod length $a = 0.5$, initial data corresponding to \eqref{61}. Simulation data  in blue, exact solution in red.}
\label{f4}
\end{figure}
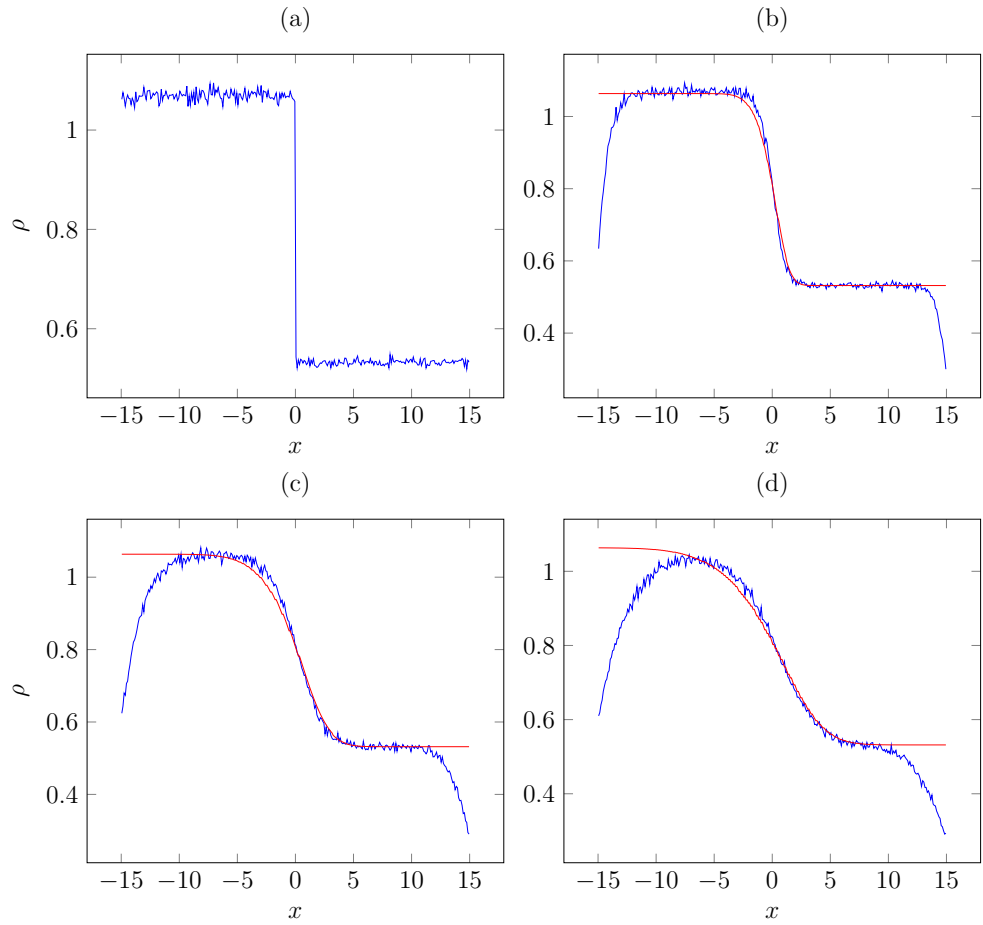

{\ }
\medskip\\
{\bf 1.} Rod length $a=0.001$, initial condition:
\beq \label{59}
	f(x,0;v) = \lt\{\ba{ll} 100 e^{-v^2} & (x\in[-15,0), \;v\in[-10,10]) \\
	50 e^{-v^2} & (x\in[0,15], \;v\in[-10,10])\\
	0 & \mbox{(otherwise)}. \ea\rt.
\eeq
In this initial condition, the value of $a\rho\in[0,1]$ is approximately $0.18$ on the left and $0.09$ on the right, which corresponds to $\langle N\rangle = 4050$. The fluid cells are taken of length $0.1$, and an average over 200 samples is done. Excellent agreement is found for both density and current,  see Figs \ref{f2} and \ref{f2c}. Note that even after seemingly small times, the system is well into the Euler scale, because the rod length and the rod separations are very small as compared to the scale at which the graph is displayed.\medskip\\
{\bf 2.} Rod length $a=0.001$, initial condition:
\beq \label{60}
	f(x,0;v) = \lt\{\ba{ll} 200 e^{-v^2} & (x\in[-5,0), \;v\in[-10,10]) \\
	75 e^{-v^2} & (x\in[0,5], \;v\in[-10,10])\\
	0 & \mbox{(otherwise)}. \ea\rt.
\eeq
In this initial condition, the value of $a\rho$ is approximately $0.36$ on the left and $0.12$ on the right, which corresponds to $\langle N\rangle = 2400$. Thereby one  probes more  deeply into the hard rod dynamics, further away from the free-particle limit. The fluid cells are again taken of length $0.1$, and an average over 50 samples is done. See Fig \ref{f3}. Since the segments of non-zero density are shorter, the depletion zone reaches the central region at an earlier time when compared to {\bf 1}.\medskip\\
{\bf 3.} In order to see deviations from  the exact solution at Euler scale, we investigate the case of a rod length $a=0.5$, with initial condition:
\beq \label{61}
	f(x,0;v) = \lt\{\ba{ll} 0.6 e^{-v^2} & (x\in[-15,0), \;v\in[-10,10]) \\
	0.3 e^{-v^2} & (x\in[0,15], \;v\in[-10,10])\\
	0 & \mbox{(otherwise)}. \ea\rt.
\eeq
In this initial condition, the value of $a\rho$ is approximately $0.55$ on the left and $0.25$ on the right. On average there are 25 rods in the fluid with an inter-particle distance of the order of unity. The fluid cells are again taken of length $0.1$, but there are so few rods that it is essential to average over a large number of samples, 120000 in our simulation. See Fig \ref{f4}. We see observe a slight departure from the exact solution. It is surprising that, even with so few rods, the Euler-scale solution is still fairly accurate.

\end{document}